\documentclass[twocolumn,showpacs,preprintnumbers,amsmath,amssymb]{revtex4}
\usepackage{graphicx}
\usepackage{dcolumn}
\usepackage{bm}

\begin{document}


\title{Chaos driven fusion enhancement factor at astrophysical energies}
\author{Sachie Kimura}
\author{Aldo Bonasera}
\affiliation{Laboratorio Nazionale del Sud, INFN,
via Santa Sofia, 62, 95123 Catania, Italy}
\date{\today}


\begin{abstract}
We perform molecular dynamics simulations 
of screening by bound target electrons in low energy nuclear reactions.
Quantum effects corresponding to the Pauli and Heisenberg principle are 
enforced by constraints. We show that 
the enhancement of the average cross section and of its variance is due to 
the perturbations induced by the electrons.
This gives a 
correlation between the maximum amplitudes of the inter-nuclear oscillational motion 
and the enhancement factor. It 
suggests that the chaotic behavior of the electronic motion affects the magnitude 
of the enhancement factor.

\end{abstract}


\maketitle

The knowledge of the bare nuclear reaction rates at low energies is
essential not only for the understanding of various astrophysical nuclear
problems, but also for assessing the effects of the host materials in low energy nuclear 
fusion reactions in matter.  This is currently a subject of great 
interest in nuclear physics, since Muenster group has reported that the
experimental cross sections of the $^{3}$He(d,p)$^{4}$He and of $^2$H($^{3}$He,p)$^{4}$He 
reactions with gas targets show an increasing enhancement with
decreasing bombarding energy with respect to the values obtained by
extrapolating from the data at high energies~\cite{krauss}. 
Many studies attempted to attribute the enhancement of the reaction
rate to the screening effects by bound target electrons.
In this context one often estimates the 
screening potential as a constant decrease of the barrier height in the tunneling region through 
a fit to the data. A puzzle has been that the screening potential obtained by this procedure 
exceeds the value of the so called adiabatic limit, 
which is given by the 
difference of the binding energies of the united atoms and of the target atom
and it is theoretically thought to provide the maximum screening potential~\cite{rolfs95}.
Over these several years, the redetermination of the bare cross sections has been proposed 
theoretically~\cite{barker} and experimentally~\cite{junker}, using the Trojan Horse 
Method~\cite{thm}.
The comparison between newly obtained bare cross sections, i.e., astrophysical 
S-factors, and the cross sections by the direct measurements gives a variety of values 
for the screening potential. 
There are already some theoretical studies performed 
using the time-dependent Hartree-Fock(TDHF) scheme~\cite{skls,ktab}.  

In this letter we examine the subject within the constrained molecular dynamics (CoMD) 
model~\cite{pmb},
even in the very low incident energy region not reached experimentally yet. At such very 
low energies fluctuations are anticipated to play a substantial role. 
Such fluctuations are beyond the TDHF scheme.
Not only TDHF calculations are, by construction, cylindrically symmetric around the beam axis. 
Such a limitation is not necessarily true in nature and the mean field dynamics could be not 
correct especially in presence of large fluctuations.     
Molecular dynamics contains all possible correlations and fluctuations due to 
the initial conditions(events). For the purpose of treating quantum-mechanical systems 
like target atoms and molecules, we use classical equations of motion with
constraints to satisfy the Heisenberg uncertainty principle and the Pauli exclusion 
principle for each event~\cite{pmb}. 
In extending the study to the lower incident energies, we would like to stress the 
connection between the motion of bound electrons and chaos. 
In fact, depending on the dynamics, the behavior of 
the electron(s) is unstable and influence the relative motion
of the projectile and the target.  
We could compare the D+d case to the gravitational 3-body problem, which 
has the same form of the equation of motion and it is nonintegrable~\cite{schuster}.
For instance, the motion of asteroids around the sun perturbed by 
Jupiter becomes unstable, i.e., chaotic, 
depending on the ratio of the unperturbed frequencies of the asteroid and Jupiter.   
We discuss the enhancement factor of the laboratory cross section in connection with 
the integrability of the
system by looking the inter-nuclear and electronic oscillational motion.
More specifically we analyze the frequency shift of 
the target electron due to the projectile and the small oscillational motion induced 
by the electron to the relative motion between the target and the projectile. 
We show that the increase of chaoticity in the electron motion decreases the fusion probability.   
In this letter we will discuss the D+d case only because the system is fundamental to see 
its connection with chaos and has been well studied theoretically. We mention that the 
understanding of the fusion dynamics and fluctuations has a great potential for 
the enhancement of the fusion probability in plasmas for energy production.

We denote the reaction cross section at incident energy in the center of mass $E$ 
by $\sigma(E)$ and the cross section obtained in absence of electrons by $\sigma_0(E)$.
The enhancement factor $f_e$ is defined as 
\begin{equation}
  \label{eq:fenh}
  f_e\equiv\frac{\sigma(E)}{\sigma_0(E)}.
\end{equation}
If the effect of the electrons is well represented by the constant shift 
$U_e$ of the potential barrier, following~\cite{alr,skls}, 
($U_e \ll E$):
\begin{equation}
  \label{eq:fenh3}
  f_e\sim\exp\left[\pi\eta(E)\frac{U_e}{E}\right],
\end{equation}
where $\eta(E)$ is the Sommerfeld parameter~\cite{clayton}.

We estimate the enhancement factor $f_e$ numerically using molecular dynamics approach;
\begin{equation}
  \label{eq:rt}  
  \frac{d {\bf r}_i}{dt}= \frac{{\bf p}_i}{{\mathcal E}_i}, \hspace*{0.5cm} 
  \frac{d {\bf p}_i}{dt}= -\nabla_{{\bf r}} U({\bf r}_i),
\end{equation}
where ({\bf r}$_i$,{\bf p}$_i$) are the position, momentum of the particle $i$
at time $t$.    
${\mathcal E}_i=\sqrt{{\bf p}_i^2c^2+m_i^2c^4}$, 
$U({\bf r}_i)$ and $m_i$ are its energy, Coulomb potential and mass, respectively. 
We set the starting point of the reaction at 10\AA~inter-nuclear separation. 
Initially the electron is located in a Bohr orbit. 
To take into account the quantum mechanical feature of atoms, we put the 
constraints, i.e., Heisenberg uncertainty principle 
and Pauli principle for atoms which have more than 2 bound electrons. 
It is performed numerically by checking $\Delta {\bf r}\cdot \Delta {\bf p}\sim\hbar$, 
for Heisenberg principle, and 
$\Delta {\bf r}\cdot \Delta {\bf p}\sim2\pi\hbar(3/4\pi)^{2/3}$, for Pauli blocking.
Here $\Delta {\bf r}=|{\bf r}_i-{\bf r}_j|$ and 
$\Delta {\bf p}=|{\bf p}_i-{\bf p}_j|$. $i$ and $j$ refer to electrons and nuclei.
More specifically, to get the atomic ground states, at every time step of the calculation, 
we calculate 
$\Delta {\bf r}\cdot \Delta {\bf p}$ for every pair of particles. 
If $\Delta {\bf r}\cdot \Delta {\bf p}$ is smaller(larger) than $\hbar$, in the case 
of Heisenberg principle, we change ${\bf r}_j$ and ${\bf p}_j$ slightly, so that
$\Delta {\bf r}\cdot \Delta {\bf p}$ becomes larger(smaller) at the subsequent time step.
We repeat this procedure for many time steps until no changes are seen in the energies and 
mean square radii of the atoms, similarly for the Pauli principle. 
The approach has been successfully applied to treat fermionic properties of 
the nucleons in nuclei and the quark system~\cite{pmb}. It can be extended 
easily in the case of the Heisenberg principle, as stated above. 
In this way we obtain many initial conditions which occupy different points in the phase 
space microscopically. 
Notice that in the D+d case that we investigate here, since one electron is involved, only 
the Heisenberg principle is enforced for each event. We obtain $-13.56$ eV as the binding energy of 
deuterium atom and 0.5327 \AA ~as its mean square radius. These values can be compared 
with the experimental value of $-13.59811$ eV \cite{bb} 
and Bohr radius $R_B=0.529$ \AA, respectively~\cite{kb-2}.   

In order to treat the tunneling process, we define the collective coordinates ${\bf R}^{coll}$ and 
the collective momentum ${\bf P}^{coll}$ as
\begin{equation}
  {\bf R}^{coll} \equiv {\bf r}_P-{\bf r}_T;   \hspace*{0.5cm}
  {\bf P}^{coll} \equiv {\bf p}_P-{\bf p}_T, 
\end{equation}
where ${\bf r}_T, {\bf r}_P$ ($ {\bf p}_T, {\bf p}_P$) are the coordinates(momenta) of the target 
and the projectile nuclei, respectively. 
When the collective momentum becomes zero, we switch on 
the collective force, which is determined by ${\bf F}_P^{coll} \equiv \dot{\bf P}^{coll}$ and 
${\bf F}_T^{coll} \equiv -\dot{\bf P}^{coll}$, to enter into imaginary time~\cite{bk}.
We follow the time evolution in the tunneling region using the equations,
\begin{equation}
  \label{eq:rti}  
  \frac{d {\bf r}_{T(P)}^{\Im}}{d\tau}= \frac{{\bf p}_{T(P)}^{\Im}}{{\mathcal E}_{T(P)}}; \hspace*{0.5cm} 
  \frac{d {\bf p}_{T(P)}^{\Im}}{d\tau}= -\nabla_{{\bf r}} U({\bf r}_{T(P)}^{\Im}) -2{\bf F}^{coll}_{T(P)},
\end{equation}
where $\tau$ is used for imaginary time to be distinguished from real time $t$.   
${\bf r}^{\Im}_{T(P)}$ and ${\bf p}^{\Im}_{T(P)}$ are position and momentum of the target
(the projectile) during the tunneling process respectively.   
Adding the collective force corresponds to inverting the potential barrier
which becomes attractive in the imaginary times.    
The penetrability of the barrier is given by~\cite{bk} 
\begin{equation}
  \label{eq:penet}
  \Pi(E)=\left(1+\exp\left(2{\mathcal A}(E)/\hbar\right)\right)^{-1},
\end{equation}
where the action integral ${\mathcal A}(E)$ is 
\begin{equation}
  {\mathcal A}(E)=\int_{r_b}^{r_a}{\bf P}^{coll}~d{\bf R}^{coll},  
\end{equation}
$r_a$ and $r_b$ are the classical turning points. The internal classical turning point 
$r_b$ is determined using the sum of the radii of the target and projectile nuclei.
Similarly from the simulation without electron, we obtain the penetrability of the bare 
Coulomb barrier $\Pi_0(E)$.

Since nuclear reaction occurs with small impact parameters on the atomic scale,
we consider only head on collisions. 
The enhancement factor is thus given by eq.\eqref{eq:fenh}, 
\begin{equation}
  f_e=\Pi(E)/\Pi_0(E)
\end{equation}
for each event in our simulation. 
Thus we have an ensemble of $f_e$ values at each incident energy. 

\begin{figure}[htbp]
  \centering
  \includegraphics[width=8.3cm,clip]{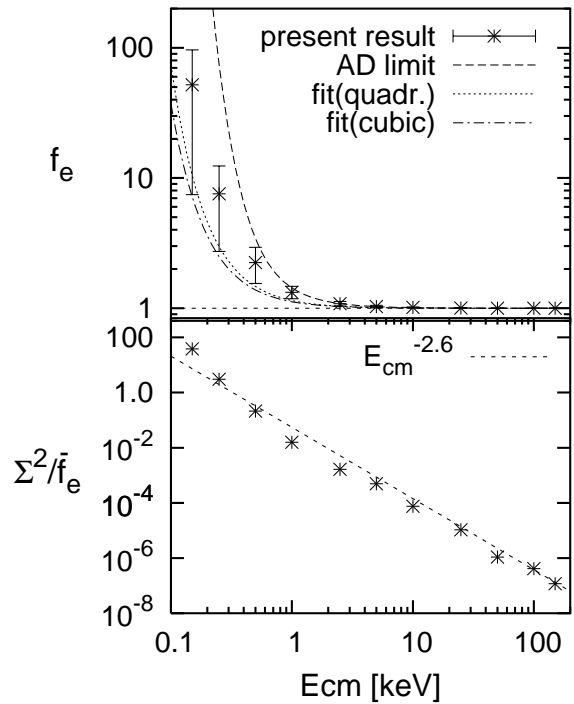}
  \caption{Enhancement factor as a function of incident center-of-mass energy for the D+d 
    reaction (upper panel).
    The corresponding $\Sigma^2/\bar{f_e}$(stars) and a power low fit(dashed line) (lower panel).}
  \label{fig:ef}
\end{figure}

In Fig.~\ref{fig:ef}, the upper panel
shows the incident energy dependence of the enhancement factor for the D+d reaction.
The averaged enhancement factors $\bar{f_e}$ over events in our simulation are shown with stars 
and its variance $\Sigma=\left(\bar{f_e^2}- (\bar{f_e})^2 \right)^{1/2}$ with error bars. 
In the figure we show also several estimations of the enhancement factor by the latest 
analysis of the experimental data using quadratic(dotted) and cubic(dot-dashed) polynomial 
fitting~\cite{barker} with the screening potentials $U_e=$ 8.7 eV and 7.3 eV respectively.  
The dashed curve shows the enhancement factor
in the adiabatic limit $f_e^{(AD)}$ for an atomic deuterium target
and it is obtained by assuming equally weighted linear combination of the lowest-energy gerade 
and ungerade wave function for the electron, reflecting the symmetry in the D+d, i.e., \\
$  f_e^{(AD)}=\frac{1}{2}\left(
    \exp(\pi\eta(E)\frac{U_e^{(g)}}{E})+\exp(\pi\eta(E)\frac{U_e^{(u)}}{E})\right),$
where $U_e^{(g)}=$ 40.7 eV and $U_e^{(u)}=$ 0.0 eV \cite{ktab,skls}. 
In the low energy region the enhancement factor is 
more than 50. However the averaged enhancement factor does not exceed the adiabatic limit. 
We performed also a fit of our data using eq.~\eqref{eq:fenh3} 
and obtained $U_e=$ 15.9 $\pm$ 2.0 eV. This value, between the sudden and the adiabatic limit, 
is in good agreement with TDHF calculations\cite{skls,ktab}.

The ratio $\Sigma^2/\bar{f_e}$ versus incident energy is plotted in the lower panel of 
Fig.~\ref{fig:ef}. The numerical results(stars) display a self similar behavior which is 
well fitted by a power law with 
exponent $-2.6$. In the high energy limit the ratio approaches zero, i.e., the $f_e$ distribution 
becomes a $\delta$-function ($\Sigma=0$) and $\bar{f_e} \rightarrow 1$: no effects due
to the electronic motion. In the low energy limit $\Sigma^2/\bar{f_e} \gg 1$, which implies 
a very sensitive dependence of the dynamics on the initial conditions, i.e., 
occurrence of chaos. Thus it is the motion of the electron which sensitively couples 
to the relative motion of the ions.          

\begin{figure*}
  \centering
  \includegraphics[width=15cm,clip]{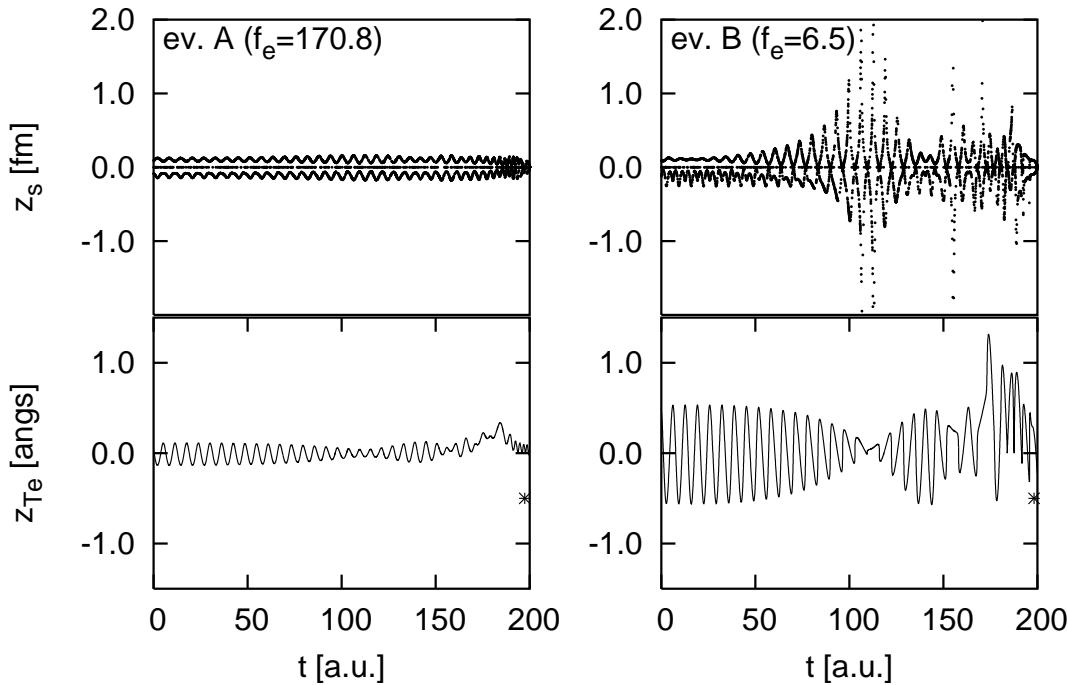}
  \caption{The oscillational motion of the electron around the target (lower panels)  
    and the inter-nuclear motion (upper panels) 
    as a function of time, in atomic unit, for two events, with large $f_e$(ev. A) 
    and small $f_e$(ev. B), for the D+d reaction at the incident 
    energy 0.15keV. The inter-nuclear separation is 10\AA~at $t=0$.}
  \label{fig:Rt}
\end{figure*}

Similar to 
the gravitational 3-body problem, we look at the oscillational motions of 
the particle's coordinates as the projection on the $z$-axis (the reaction axis). 
We denote the $z$-component of ${\bf r}_T, {\bf r}_P$ and ${\bf r}_e$ as 
$z_T, z_P$ and $z_e$, respectively.
Practically, we examine the oscillational motion of the electron around the target
$z_{Te}=z_e-z_T$ and 
the oscillational motion of the inter-nuclear motion, i.e., the motion between 
the target and the projectile, $z_{s}=z_T+z_P$, which 
essentially would be zero due to the symmetry of the system, if there were no perturbations. 
In Fig.~\ref{fig:Rt} these two values are shown for 2 events, which have the enhancement 
factor $f_e=$ 170.8 (ev. A), and $f_e=$ 6.5 (ev. B), at the incident energy 
$E_{cm}=0.15$ keV.  
The panels show the $z_s, z_{Te}$ 
as a function of time. The stars indicate the time at which the system reaches the classical turning point.
It is clear that in the case of ev. B the orbit of the electron is much distorted from 
the unperturbed one than in ev. A.
Characteristics of $z_s$ are that (1) its 
value often becomes zero, as it is expected in the un-perturbed system, and 
(2) the component of the deviation 
from zero shows periodical behavior. It is remarkable that the amplitude of the deviation becomes 
quite large at some points in the case of ev. B which shows the small enhancement factor. 
Note that in event B one observes clear beats, i.e., resonances.   
Thus for two events, with the same macroscopic initial conditions, we have a completely different 
outcome, which is a definite proof of chaos in our 3-body system. 
We can understand these results in first 
approximation by considering the motion of the ions to be 
much slower than 
the rapidly oscillating motion of the electrons. Thus we can consider
the electron acting as an external force $F_e=F_0\cos(\omega_\textrm{H} t+\delta)$, 
where $F_0$ is the amplitude of the force and $\omega_\textrm{H}$ is the (hydrogen) frequency. 
This will induce~\cite{lp} a perturbation on 
$z_s \sim F_e/{\mu \omega_\textrm{H}^2}$,
where $\mu$ is the ions reduced masses. Notice 
how the amplitude of $z_s$ is actually reduced from the amplitude of $z_{Te}$ of a factor 1 $\sim$ 
$10^{-4}$, i.e., the ratio of the electron to the ion mass. We stress that this simple estimation 
is more relevant for case A.  In fact the motion of the electron 
is not decoupled from the inter-nuclear motion 
and from energy conservation we can expect that when $|z_s|$ is maximum, $|z_{Te}|$ is minimum 
as observed in Fig.\ref{fig:Rt}, case B. Thus more generally, one should consider a perturbation 
where  $F_0\rightarrow F_0(t)$ and $\omega_\textrm{H} \rightarrow \omega_\textrm{H}(t)$.
The time dependence of the perturbation leads, as it is well known, to the occurrence of chaos 
as for parametric resonance~\cite{lp}.  From the Fig.\ref{fig:Rt} we can deduce the following 
important fact. If the motion of the electron is initially in the plane perpendicular to the reaction 
axis, the enhancement factor is large, case A(notice $|z_{Te}| \ll R_B$, i.e., the Bohr radius, at $t\sim 0$). 
On the other hand if there is a substantial projection 
of the electron motion, as in case B(the amplitude of $|z_{Te}|\sim R_B$ at $t\sim 0$), on the reaction 
axis the enhancement factor is relatively small
because of the increase of chaoticity. This suggests that if one performs experiments at very low 
bombarding energies with {\it polarized targets}, the enhancement factor can be controlled by 
changing 
the {\it polarization}. The largest enhancement with targets polarized perpendicularly to the beam 
direction.             
We notice in passing that event A is a case where cylindrically symmetry is approximately satisfied, because 
the electron motion is practically on the $xy-$plane. This case gives a $U_e=$19.5eV closest to the 
adiabatic limit and to the TDHF result\cite{skls,ktab}.  
      
\begin{figure}[htbp]
  \centering
  \includegraphics[width=7.8cm,clip]{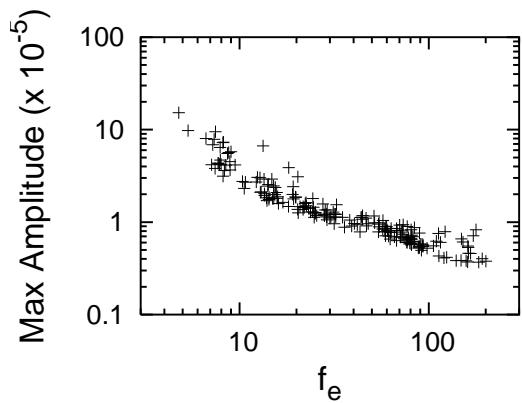}
  \caption{The maximum amplitude of $z_s$ for 200 events as a function of the enhancement 
    factor $f_e$ for the D+d reaction at the incident energy 0.15 keV. }
  \label{fig:maxamp}
\end{figure}

In Fig.~\ref{fig:maxamp} we show the maximum amplitude of $z_s$ for 200 events as a function of the 
enhancement factor $f_e$ at the incident energy $E_{cm}=0.15$ keV. 
Here we observe an evident correlation between the two values: the events which give
relatively large enhancement factors correspond to the small maximum amplitude of $z_s$, and to the 
contrary, the events with relatively small enhancement factors indicate large maximum amplitudes. 
Since the motion of the ions is coupled to that of the electrons, it implies that an amount 
of energy is transferred from the relative motion of the ions to the electron thus reducing 
the probability of fusion. 
On the contrary small amplitudes imply that the electronic configuration is 
closer to the one in the g.s. of the compound system. 
Thus the binding energy of the 
electron is converted into the relative energy of the ions increasing the fusion probabilities.

\smallskip
In conclusion, we 
discussed the penetrability of the Coulomb barrier by using 
molecular dynamics simulations with constraints and imaginary time. 
We have shown that both the enhancement factor and its variance increase as the incident energy 
becomes lower. However we obtained the averaged screening potential 
smaller than the value in the adiabatic limit,
while from fluctuations some events clearly exceed such a limit.  
We pointed out that there is an evident correlation between the oscillational motion of the 
inter-nuclear separation and the magnitude of the enhancement factor of the cross section.
The chaoticity of the electron motion affects the enhancement factor of the cross section.     
We suggest to perform experiments on fusion at very low energies with polarized targets.

We acknowledge valuable discussions and suggestions with Profs. D.M. Brink, G. Fiorentini, 
C. Rolfs, C. Spitaleri and N. Takigawa. 


\end{document}